\documentstyle[11pt,aaspp4,flushrt]{article}
\slugcomment{submitted to the Astronomical Journal}
\def\arcsec{\ifmmode '' \else $''$\fi}
\def\arcmin{\ifmmode ' \else $'$\fi}
\def\arcsecpoint{\ifmmode ''\!. \else $''\!.$\fi}
\def\arcminpoint{\ifmmode '\!. \else $'\!.$\fi}
\def\cc{\ifmmode {\rm cm}^{-3} \else cm$^{-3}$\fi}
\def\cl{\ifmmode {\rm cm}^{-2} \else cm$^{-2}$\fi}
\def\micron{\ifmmode \mu{\rm m} \else $\mu$m\fi}
\def\kms{\ifmmode {\rm km\,s}^{-1} \else km\,s$^{-1}$\fi}
\def\Hubble{\ifmmode {\rm km\,s}^{-1}\,{\rm Mpc}^{-1}
	\else km\,s$^{-1}$\,Mpc$^{-1}$\fi}
\def\ergsec{\ifmmode {\rm ergs\;s}^{-1} \else ergs s$^{-1}$\fi}
\def\ergscm{\ifmmode {\rm ergs\,s}^{-1}\,{\rm cm}^{-2}
	  \else ergs\,s$^{-1}$\,cm$^{-2}$\fi}
\def\ergscmA{\ifmmode {\rm ergs\,s}^{-1}\,{\rm cm}^{-2}\,{\rm \AA}^{-1}
	  \else ergs\,s$^{-1}$\,cm$^{-2}$\,\AA$^{-1}$\fi}
\def\ergscmHz{\ifmmode {\rm ergs\,s}^{-1}\,{\rm cm}^{-2}\,{\rm Hz}^{-1}
	  \else ergs\,s$^{-1}$\,cm$^{-2}$\,Hz$^{-1}$\fi}
\def\Msun{\ifmmode M_{\odot} \else $M_{\odot}$\fi}
\def\Lsun{\ifmmode L_{\odot} \else $L_{\odot}$\fi}
\def\qo{\ifmmode q_{0} \else $q_{0}$\fi}
\def\Ho{\ifmmode H_{0} \else $H_{0}$\fi}
\def\ltsim{\raisebox{-.5ex}{$\;\stackrel{<}{\sim}\;$}}
\def\gtsim{\raisebox{-.5ex}{$\;\stackrel{>}{\sim}\;$}}
\def\lya{Ly$\alpha$}
\def\lyb{Ly$\beta$}
\def\civ{C\,{\sc iv}}

\newcommand {\etal}{et al.}
\newcommand{\siv}{Si{\sc iv}+O{\sc iv}]}
\newcommand{\nv}{N{\sc v}}
\newcommand{\ovi}{O{\sc vi}}
\newcommand{\oi}{O{\sc i}}

\begin{document}

\title{Five High-Redshift Quasars Discovered in Commissioning Imaging Data
of
the Sloan Digital Sky Survey$^1$}

\author{Wei Zheng\altaffilmark{\ref{JHU},\ref{IRTF}},
Zlatan I. Tsvetanov\altaffilmark{\ref{JHU},\ref{IRTF}},
Donald P. Schneider\altaffilmark{\ref{PennState}},
Xiaohui Fan\altaffilmark{\ref{Princeton}},
Robert H. Becker\altaffilmark{\ref{UCD}},
Marc Davis\altaffilmark{\ref{UCB}},
Richard L. White\altaffilmark{\ref{STScI}},
Michael A. Strauss\altaffilmark{\ref{Princeton}},
James Annis\altaffilmark{\ref{Fermilab}},
Neta A. Bahcall\altaffilmark{\ref{Princeton}},
A. J. Connolly\altaffilmark{\ref{Pittsburgh}},
Istv\'an Csabai\altaffilmark{\ref{JHU},\ref{Eotvos}},
Arthur F. Davidsen\altaffilmark{\ref{JHU}},
Masataka Fukugita\altaffilmark{\ref{CosmicRay},\ref{IAS}},
James E. Gunn\altaffilmark{\ref{Princeton}},
Timothy M. Heckman\altaffilmark{\ref{JHU}},
G. S. Hennessy\altaffilmark{\ref{USNO}},
\v{Z}eljko Ivezi\'{c}\altaffilmark{\ref{Princeton}},
G. R. Knapp\altaffilmark{\ref{Princeton}},
Eric Peng\altaffilmark{\ref{JHU}},
Alexander S. Szalay\altaffilmark{\ref{JHU}},
Aniruddha R. Thakar\altaffilmark{\ref{JHU}},
Brian Yanny\altaffilmark{\ref{Fermilab}},
and Donald G. York\altaffilmark{\ref{Chicago}}
}

\altaffiltext{1}{Based on observations obtained with the
Sloan Digital Sky Survey, and with the Apache Point Observatory
3.5-meter telescope, which is owned and operated by the Astrophysical
Research Consortium, with the Hobby-Eberly Telescope (HET), which is a
joint
project of the University of Texas at Austin, the Pennsylvania State
University, Stanford University,
Ludwig-Maximillians-Universit\"at M\"unchen, and
George-August-Universit\"at
G\"ottingen, and at the W.M. Keck Observatory, which is operated as a
scientific partnership among the University of California, the California
Institute of Technology, and the National Aeronautics and Space
Administration, and was made possible by the generous financial support of
the
W.M. Keck Foundation.}

\newcounter{address}

\setcounter{address}{2}
\altaffiltext{\theaddress}{
Department of Physics and Astronomy, The Johns Hopkins University,
Baltimore, MD 21218
\label{JHU}}
\addtocounter{address}{1}
\altaffiltext{\theaddress}{Visiting Astronomer at the NASA Infrared
Telescope
Facility, which is operated by the University of Hawaii under contract
from
the National Aeronautics and Space Administration
\label{IRTF}}
\addtocounter{address}{1}
\altaffiltext{\theaddress}{Department of Astronomy and Astrophysics,
The Pennsylvania State University, University Park, PA 16802
\label{PennState}}
\addtocounter{address}{1}
\altaffiltext{\theaddress}{Princeton University Observatory, Princeton, NJ
 08544
\label{Princeton}}
\addtocounter{address}{1}
\altaffiltext{\theaddress}{Department of Physics, University of
California,
Davis, CA 95616
\label{UCD}}
\addtocounter{address}{1}
\altaffiltext{\theaddress}{Department of Astronomy, University of
California,
Berkeley, CA 94720-3411
\label{UCB}}
\addtocounter{address}{1}
\altaffiltext{\theaddress}{
Space Telescope Science Institute, Baltimore, Maryland 21218
\label{STScI}}
\addtocounter{address}{1}
\altaffiltext{\theaddress}{Fermi National Accelerator Laboratory, P.O.
 Box 500, Batavia, IL 60510
\label{Fermilab}}
\addtocounter{address}{1}
\altaffiltext{\theaddress}{Department of Physics and Astronomy,
University of Pittsburg, Pittsburgh PA 15260
\label{Pittsburgh}}
\addtocounter{address}{1}
\altaffiltext{\theaddress}{Department of Physics of Complex Systems,
E\"otv\"os University,
P\'azm\'any P\'eter s\'et\'any 1/A, Budapest, H-1117, Hungary
\label{Eotvos}
}
\addtocounter{address}{1}
\altaffiltext{\theaddress}{Institute for Cosmic Ray Research, University
of
Tokyo, Midori, Tanashi, Tokyo 188-8502, Japan
\label{CosmicRay}}
\addtocounter{address}{1}
\altaffiltext{\theaddress}{Institute for Advanced Study, Olden Lane,
Princeton, NJ 08540
\label{IAS}}
\addtocounter{address}{1}
\altaffiltext{\theaddress}{U.S. Naval Observatory,
3450 Massachusetts Ave., NW, Washington, DC  20392-5420
\label{USNO}}
\addtocounter{address}{1}
\altaffiltext{\theaddress}{University of Chicago, Astronomy \&
Astrophysics
Center, 5640 S. Ellis Ave., Chicago, IL 60637
\label{Chicago}}

\begin{abstract}
We report the discovery of five quasars with redshifts of $4.67 - 5.27$
and $z'$-band magnitudes of $19.5-20.7$ ($M_B \sim -27$).
All were originally selected as distant quasar candidates in
optical/near-infrared photometry from the Sloan Digital Sky Survey (SDSS),
and most were confirmed  as probable high-redshift quasars
by supplementing the SDSS data with $J$ and $K$ measurements.
The quasars possess strong, broad \lya\  emission lines, with
the characteristic sharp cutoff on the blue side produced by
\lya\ forest absorption. Three quasars contain strong, broad absorption
features, and one of them exhibits very strong \nv\ emission.
The amount of absorption produced
by the \lya\ forest increases toward higher redshift, and that in the
$z$~=~5.27 object \hbox{($D_A \approx 0.7$)}
is consistent with a smooth extrapolation of the absorption seen in lower
redshift quasars. The high luminosity of these objects relative
to most other known objects at $z \gtsim 5$ makes them potentially
valuable as
probes of early quasar properties and of the intervening intergalactic
medium.
\end{abstract}

\section{Introduction}

Since the identification of the first quasar redshift
(3C 273, \cite{schmidt63}), quasars have been
at the forefront of modern cosmology. With luminosities tens or hundreds
of
times higher than those of galaxies, quasars are a powerful probe of the
distant primordial universe. Over the past decade approximately two
hundred
objects above redshift four have been discovered, and there is a growing
consensus
that the number density of luminous quasars peaks between redshifts
of two and three and steeply declines out to the limits of current
measurements
($z \approx 4.5$; see
Warren, Hewett, \& Osmer 1994; \cite{ssg95}; \cite{kddc95}). Recent
studies (\cite{fan99},2000) have dramatically increased the number of
known quasars with redshifts larger than~4.5, opening the possibility
of investigating the quasar luminosity function at redshift five and
beyond; this information will determine whether the number density
of quasars continues to decline with increasing redshift, or whether
theoretical models ($e.g.,$ \cite{hl98}) that
predict a significant number of quasars at $ z > 5$ are correct.

The majority of high-redshift quasars have been identified by optical
color
selection. As the result of intergalactic absorption, the flux in the
spectral region shortward of \lya\ in distant objects is
significantly attenuated. The onset of the \lya\ forest
and the Lyman break can be detected with
broad-band imaging. Multicolor optical surveys (\cite{who91},
\cite{imh91},
\cite{sgd99})
have proven effective in identifying $z>4$ quasars, but as most such
surveys
lack information redward of the $I$ band, they have difficulty
distinguishing
between quasars at redshifts larger than $\approx$ 4.8 and cool stars.
Recently, the commissioning data of the Sloan Digital Sky Survey (SDSS,
\cite{york00}) have led to the identification of new high-redshift quasars
at
an unprecedented rate: approximately 45 quasars at $z> 3.6$ (two at
$z \sim 5$, \cite{fan99},2000; \cite{HET}), have been published in the
past two years. These quasars, at $M_B\sim
-27$, are at the luminous end of the quasar luminosity distribution.
(Throughout this paper cosmological properties are calculated assuming
\hbox{$H_0$ = 50 km s$^{-1}$ Mpc$^{-1}$} and~$q_0$~=~0.5.)

In addition to wide area surveys, color selection for high-redshift
objects has been applied to small, deep fields.
Three galaxies at $ z > 5.6$ have been spectroscopically confirmed
(\cite{hu98}; \cite{weymann98}; \cite{hu99}), and recently
\cite{stern00} found an AGN ($M_B \sim -22.7$) at $z=5.5$.
The Hubble Deep Field has led to the discovery of
high-redshift candidates ($z \sim$ 6-10, \cite{lanzetta99};
\cite{chen99});
these objects, however,
are orders of magnitude less luminous than quasars, and their
spectral properties are difficult to study even with the world's largest
telescopes.

Although the $z=5$ barrier was broken more than two years ago, objects
at these extreme redshifts are still rare enough that each new example is
a potentially valuable probe of the very early universe.  This is
especially true in the case of the most luminous quasars, since
follow-up spectroscopy of these relatively bright objects can reveal
information about the physical conditions in early quasars
and about the state of the intergalactic medium at very high redshifts.
The SDSS collaboration has undertaken extensive efforts to search for
high-redshift quasars.
In this paper, we report the discovery of five quasars at $z>4.6$,
of which the most distant is at  $z=5.27$.

\section{Selection of Quasar Candidates}

The SDSS (York et al. 2000) utilizes a wide-field camera with 54
CCDs (\cite{gunn98}),
mounted on a dedicated 2.5m telescope at the Apache Point Observatory
(APO), New Mexico, to survey $\approx$~$\pi$ steradians
of the sky around the Northern Galactic Cap. CCD images in five broad
optical
bands ($u'$, $g'$, $r'$, $i'$, $z'$, centered at 3540~\AA, 4770~\AA,
6230~\AA,
7630~\AA\ and~9130~\AA\ ; \cite{fukugita96}) yield a nominal
$5\sigma$ detection
of point sources in AB magnitudes of 22.3, 23.3, 23.1, 22.3, and 20.8,
respectively. The commissioning data have so far covered $\sim 600$ square
degrees, mostly near the equatorial region, $|\delta| < 1.5 ^\circ$.

With five bands and a spectral resolution of $R \sim 3$, the SDSS imaging
data
can distinguish quasars from stars in a broad redshift range (\cite{fan}).
At $3.5 \ltsim z \ltsim 5$,
quasars lie well away from the stellar locus in the $g'r'i'z'$
color space, due to the large equivalent width of the \lya\ emission
line and the significant absorption produced by the \lya\ forest
and Lyman limit systems.  At redshifts between 4.4 and 5, the $r'-i'$
color
becomes large, and $i'-z'$ is near zero. At $ z \gtsim 5$, the flux drop
shortward of the redshifted \lya\ emission line affects
the $i'$-band magnitude, and the $i'-z'$ color increases with redshift.
At this point the quasar track quickly approaches
the red end of the stellar locus in the $r'i'z'$ color diagram;
the rise in contamination by very cool stars requires that additional
discriminators be added to aid the
quasar selection. For quasars, the underlying continuum longward of \lya\
can
be approximated with a power law (\cite{francis91}; \cite{ssg};
\cite{zheng97}) of $f_\nu \propto
\nu^{-0.9} $, leading to a small color differences in bands located
redward of the \lya\ emission line
\hbox{($i.e.,$ blue band $-$ red band $\approx$ 0).}
This small color difference contrasts sharply with the colors of
most cool stars (\cite{leggett96}, 2000), whose flux rises rapidly
towards longer wavelengths.

Our target selection is based on three regions in SDSS color space:
(1) $r^*-i^* > 1.35$ and $i^*-z^* < 0.3$; (2) $r^*-i^* > 2$
and $i^*-z^* <0.7$; or (3) $z'$-band detection only, i.e. $z^* <20.8$ and
the
detection in the other four bands is below $5 \sigma$. If an object's
$r'$ or $i'$ magnitude is below the respective $5\sigma$ detection level,
we
use the latter in calculating the color in case (1) and (2).
(Note the $^*$ superscript for the
magnitudes; the measurements reported here are based on a preliminary
calibrations of the SDSS photometric system.)
In addition, an object must be classified as a point
source by the SDSS processing software and have
\hbox{$z^* \le 20.8$} to be included as a quasar candidate.
We applied these selection criteria to $\sim 200$ square degrees of
SDSS imaging data acquired in 1999 March and 2000 February.

The selection criteria used here differ from those employed
by Fan et al. (1999, 2000) to identify $z > 4$ quasars. Whereas Fan et al.
required $i^* < 20.2$, candidates in this paper
can be undetected in the $i'$ band. The flux measurements for this paper's
candidates often have lower signal-to-noise ratio, and a sample drawn
from these selection criteria will naturally be more susceptible to
contamination from non-quasars than those in Fan et al.  The color cuts
in this paper are closer to the stellar locus than the Fan et al.
criteria;
this will also resultin a reduction in the selection efficiency.

We obtained IR photometry of quasar candidates in the $J$ and
$K$ bands at the NASA Infrared Telescope
Facility (IRTF) at Mauna Kea, Hawaii.
The observations were taken on 2000 March 9-12 using the NSFCam
equipped with a 256$\times$256 InSb array.
The plate scale was 0\farcs30 pixel$^{-1}$, and the seeing was
$\sim$1\arcsecpoint 2 in the $K$-band. Each selected target was imaged
with
standard dithering technique with total exposure times of 7 min and 5
min in $J$ and $K$, respectively. Images of IR standard stars were
taken throughout the observations to monitor the magnitude zero points.
Typical photometric errors are in the 0.05-0.1 magnitude range,
depending on the brightness of objects;
see \cite{zt00} for details of the observations and data calibration.
We observed $\sim 20$ known SDSS quasars with redshifts greater than 4
to empirically calibrate our selection technique.
Our additional color constraints are $z^* - J < 1.5$ and $J - K < 1.8$.
Approximately ten such objects whose colors closely resembled those of
known
quasars were selected. Note that four of the five quasars are at redshift
smaller than 5, and
they can be selected by the SDSS data alone (\cite{fan99}, 2000).


\section{Spectroscopic Observations}

Spectroscopic follow-up observations of SDSS high-redshift quasar
candidates
were carried out in 2000 February and March, with the Digital Imaging
Spectrograph (DIS) of the APO 3.5m telescope.
The DIS is a double spectrograph; for high-redshift quasars, only the
red part of the low-resolution spectrum, covering the wavelength range
5400~\AA\ to 10000~\AA\ at 13~\AA\ resolution, contained any useful
signal.
The exposure time for each object was 30 minutes; even with a limited
signal-to-noise ratio in the spectra, the redshift identifications, based
on
the strong, asymmetric \lya\ emission line and absorption produced by the
\lya\ forest, are unambiguous. The observations in February were made
before
the IRTF run. Of the six candidates, only one, SDSS~1129$-$0142, turned
out
to be a quasar. The J2000
coordinates are given in the object name (format
\hbox{hhmmss.ss+ddmmss.s;}
see \cite{fan99}.)  For brevity, we have shortened the names to
SDSS~hhmm+ddmm throughout the text.)
In late March we observed five SDSS/IR candidates, and two,
SDSS~1021$-$0309 and SDSS~1208+0010, are quasars.

After the initial identification with the APO data, additional
spectroscopic
observations of the three quasars were obtained in 2000~April with the
Low Resolution Spectrograph (LRS; \cite{hill00}, \cite{HET}) at the prime
focus
of the Hobby-Eberly Telescope (\cite{lwr98}) at McDonald Observatory.  The
LRS configuration of a 2$''$ slit, 300~line~mm$^{-1}$ grating, and OG515
blocking filter produced spectra with a wavelength coverage of
5150~\AA\ to 10,150~\AA\ at a resolution of~20~\AA .  The exposure times
were typically~30~minutes.

Two additional SDSS/IR quasar candidates were observed with the Keck 10m
telescope on April 5-6, and both of them, SDSS~1451$-$0104 and
SDSS~1122$-$0229, are quasars.
The spectra were taken with the Echellette spectrograph and
imager (ESI, \cite{esi}) on the Keck Observatory
10-m telescope. The ESI was used in high dispersion mode
which covers the wavelength range of 4000 to 11000 \AA\ with
11 \kms\ resolution. The quasars were viewed through a 1\arcsec\
wide slit oriented at the parallactic angle. The exposure time is 20
minutes
each.
The $z=5.27$ quasar, SDSS 1208+0010, was also observed for 30 minutes.
The spectra were flux calibrated relative to the standard star G191B2B.
The
spectra were extracted and reduced using standard IRAF programs, and
binned
to 3.85\AA.

Only one of the five ``non-infrared" candidates, observed in February,
is a quasar. Of the seven SDSS/IR candidates observed, four are quasars,
while the other are late-type stars. This result tentatively suggests that
IR selection may significantly improve the selection efficiency, but
clearly
a larger sample is needed to confirm this conclusion.
Table~1 lists the optical/IR photometric measurements.

The quasar spectra are displayed in Fig. 2, with prominent spectral
features
marked. 
All the spectra have been
placed on an absolute flux scale by matching the $i^*$ magnitudes in
Table~1
with the $i^*$ magnitudes synthesized from the spectrum.
The Keck and HET spectra of SDSS~1208+0010 reveal the \civ\ emission at
$\sim
9500$\AA, which is not clear in the data taken at APO.

\section{Discussion}

Table~2 contains the redshift (see notes below for the specifics of the
measurements for each object), AB$_{1450}$ (the AB magnitude of the
quasar at 1450~\AA\ in the rest frame, corrected for Galactic absorption),
the power law index of the continuum, the continuum depression due to
absorption in the \lya\ and \lyb\ regions ($D_A$ and $D_B$;
see \cite{OK82}), and the absolute $B$ magnitude of the quasars.
The Galactic extinction is calculated using the reddening map of
\cite{Schlegel98}.
The luminosities for the quasars were calculated assuming that the
continuum
power law slope from the far ultraviolet to the optical was~$-0.5$
(\cite{ssg}). The individual
power law slopes, and hence the depression estimates, are quite uncertain
given the limited baseline available in the spectra and the
presence of BAL features. All are moderately
luminous quasars, and the spectra of three contain BAL features.  Each
spectrum contains at least one excellent candidate for a damped
\lya\ system, and there is no detected flux below the rest frame Lyman
limit
in any of the quasars.

As shown in Figure 2, the extent of intergalactic absorption shortward of
the
redshifted \lya\ emission increases from $z=4.67$ to $z=5.27$. This is
reflected in the continuum depression values in Table 2. However, the
residual
flux can be seen in all the spectra, particularly around the \ovi+\lyb\
feature, suggesting that the intergalactic medium is not completely
opaque. Our
measurements are consistent with the known distributions of the \lya\
forest absorption as derived at lower redshifts (\cite{press93}).

The emission features in these quasars are common among AGN at lower
redshift.
Three of the the quasars exhibit significant BAL features.
Finding charts for the five quasars are given in Fig.~1.

\noindent
Notes on the quasars:

\noindent
{\bf SDSS 1021$-$0309} ($z$ = 4.70): The sharp split of the \lya\
and \nv\ lines may suggest unusually strong \nv\ emission. A simple fit to
a 300\AA- region longward of the \lya\ cutoff yields a \nv/\lya\ ratio of
0.64
for the narrow features. If the split is a result of \nv\ absorption,
the corresponding wavelength for the \civ\ absorption counterpart should
be
centered at $\sim 8710$\AA. While a significant \civ\ BAL trough is
present
in the spectrum, it is centered at $\sim 8560$\AA.
The redshift determined by the
\oi, \siv, and \civ\ emission lines are consistent with each other at
the~0.005 level.
The \lya\ and \nv\ emission shows redshifts that are
consistent (with slightly larger errors) with the other lines.

\noindent
{\bf SDSS 1122$-$0229} ($z$ = 4.80): This object has the strongest
emission
lines among the five. The measurements of \civ, \oi, \nv\ and \lya\
emission
yields a redshift of $z = 4.80 \pm 0.03$. The \lya\ absorption cutoff is
not
as sharp as the others, stretching $\sim 60$\AA\ with several narrow
absorption troughs.

\noindent
{\bf SDSS 1129$-$0142} ($z$ = 4.85): This quasar is slightly more luminous
than
3C~273, which in the adopted cosmology is $M_B = -27.0$. It displays a
number
of spectacular BAL features; because of this, many of the properties given
in Table~2 contain large uncertainties.  Strong, broad troughs of
\civ\ and \siv\ 
dominate the spectrum, and there is a suggestion of
the presence of 
\oi\ and \nv\ absorption features.  The redshift is
based on assigning a rest wavelength of~1219~\AA\ to
the peak of the \lya\ emission line (see \cite{ssg}).  The observed values
of
the continuum depression are exceptionally large; this almost certainly
arises from significant intrinsic absorption.

\noindent
{\bf SDSS 1208+0010} ($z$ = 5.27): Very strong, relatively
narrow \lya\ and \nv\ emission dominate the spectrum; this spectrum
bears an uncanny resemblance to that of the $z = 4.04$ quasar
\hbox{PC 0910+5625} (\cite{ssg87}). The \nv,\oi, 
and
\civ\ lines yield a consistent redshift; the peak of the \lya\ feature
occurs
at~1218~\AA, typical for quasars at redshifts above four (\cite{ssg}).
The depression due to the \lya\ forest (see Table~2) is quite
large, but not unusual for this redshift,
suggesting that there is not a dramatic change in the character in the
\lya\ forest from lower redshifts to at least $z \sim 5.2$.

\noindent
{\bf SDSS 1451$-$0104} ($z$ = 4.67):
The redshifts determined from the peaks of \civ\ and \siv\ match that of
the
\lya\ edge within 0.005. A significant, broad absorption trough is present
between $\sim 8250 - 8550$\AA.

The discovery of these quasars once again demonstrates the ability of
the SDSS to effectively identify $z>4.6$ quasars, and extends the SDSS
redshift
range to well beyond five.  By supplementing the SDSS measurements with
$J$ and $K$ photometry, we have been able to efficiently identify
(success rate of $\approx$ 50\%) high-redshift quasars in magnitude/color
space regions that are fainter and closer to the stellar locus than were
presented in Fan et al. (1999,2000); it is likely that IR photometry
will be a valuable tool in the search for faint, $z > 5$ quasars.
Our IR photometry is only a test, and the results do not constitute a
complete sample.
To date the SDSS has imaged but a few percent of the
planned survey area; based on the results to date,
the complete survey should contain well over a hundred $z>4.7$ quasars
found with well-defined selection criteria.


\acknowledgments

The Sloan Digital Sky Survey (SDSS) is a joint project of The
University of Chicago, Fermilab, the Institute for Advanced Study, the
Japan Participation Group, The Johns Hopkins University, the
Max-Planck-Institut f\"ur Astronomie, Princeton University, the United
States Naval Observatory, and the University of Washington. Apache
Point Observatory, site of the SDSS, is operated by the Astrophysical
Research Consortium. Funding for the project has been provided by the
Alfred P. Sloan Foundation, the SDSS member institutions, the National
Aeronautics and Space Administration, the National Science Foundation,
the U.S. Department of Energy, and Monbusho, Japan. The SDSS Web site is
http://www.sdss.org/.

We would like to thank Russet McMillan (APO), William Golisch (IRTF),
Bob Goodrich, Terry McDonald (Keck),
Grant Hill and Matthew Shetrone (HET) for their assistance with
the observations.
David Weinberg provided a number of comments that improved the paper.
This work is partially supported by NASA Long Term
Space Astrophysics grant NAGW-4443 to the Johns Hopkins University (WZ and
ZT),
and by NSF grant AST99-00703 (DPS).

\newpage

\figcaption[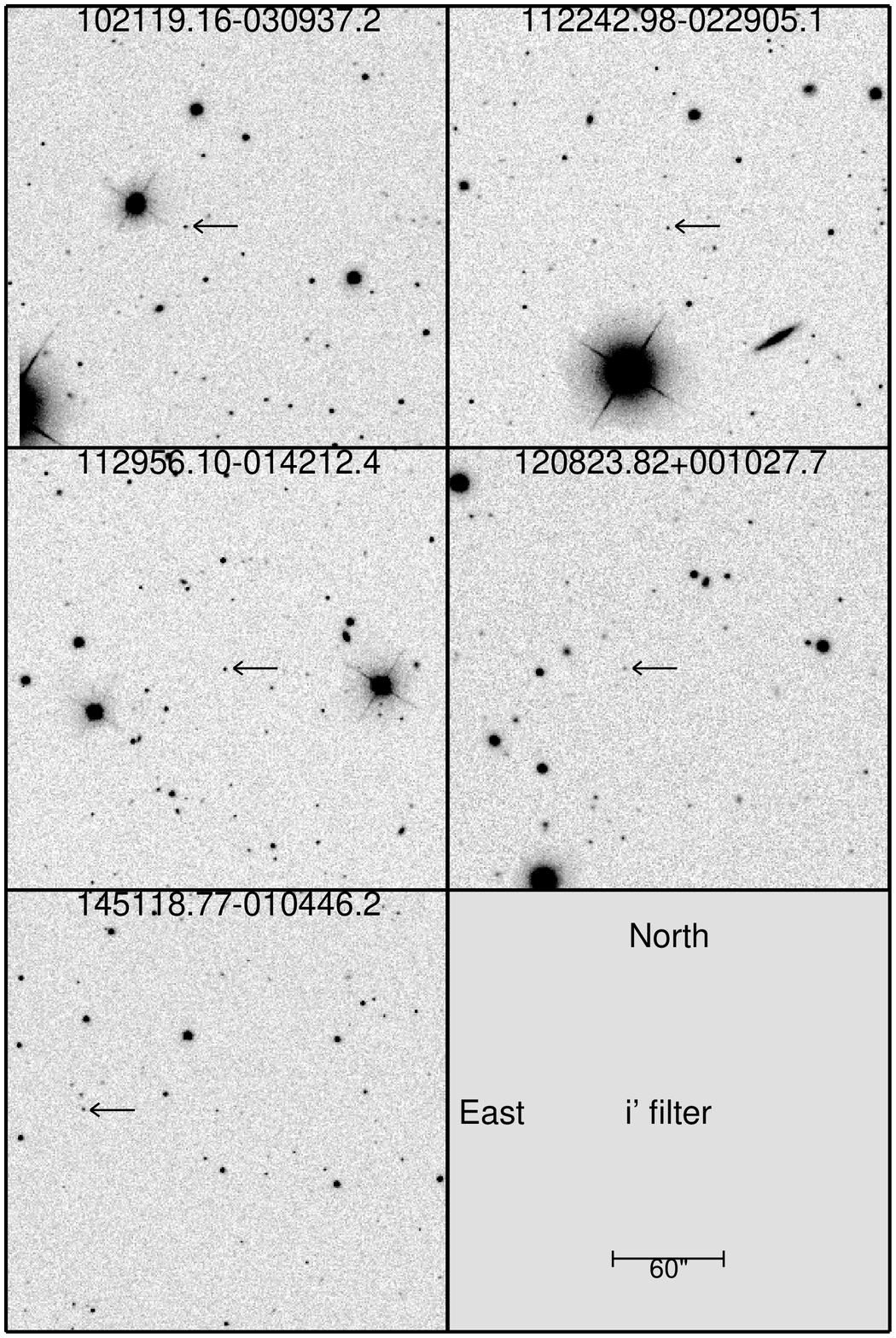]{Finding charts of the five quasars. All the
charts are
$i'$-band images from the SDSS. North is up, East to the left, and the
individual charts are 240$''$ on a side.
\label{fig1}}

\figcaption[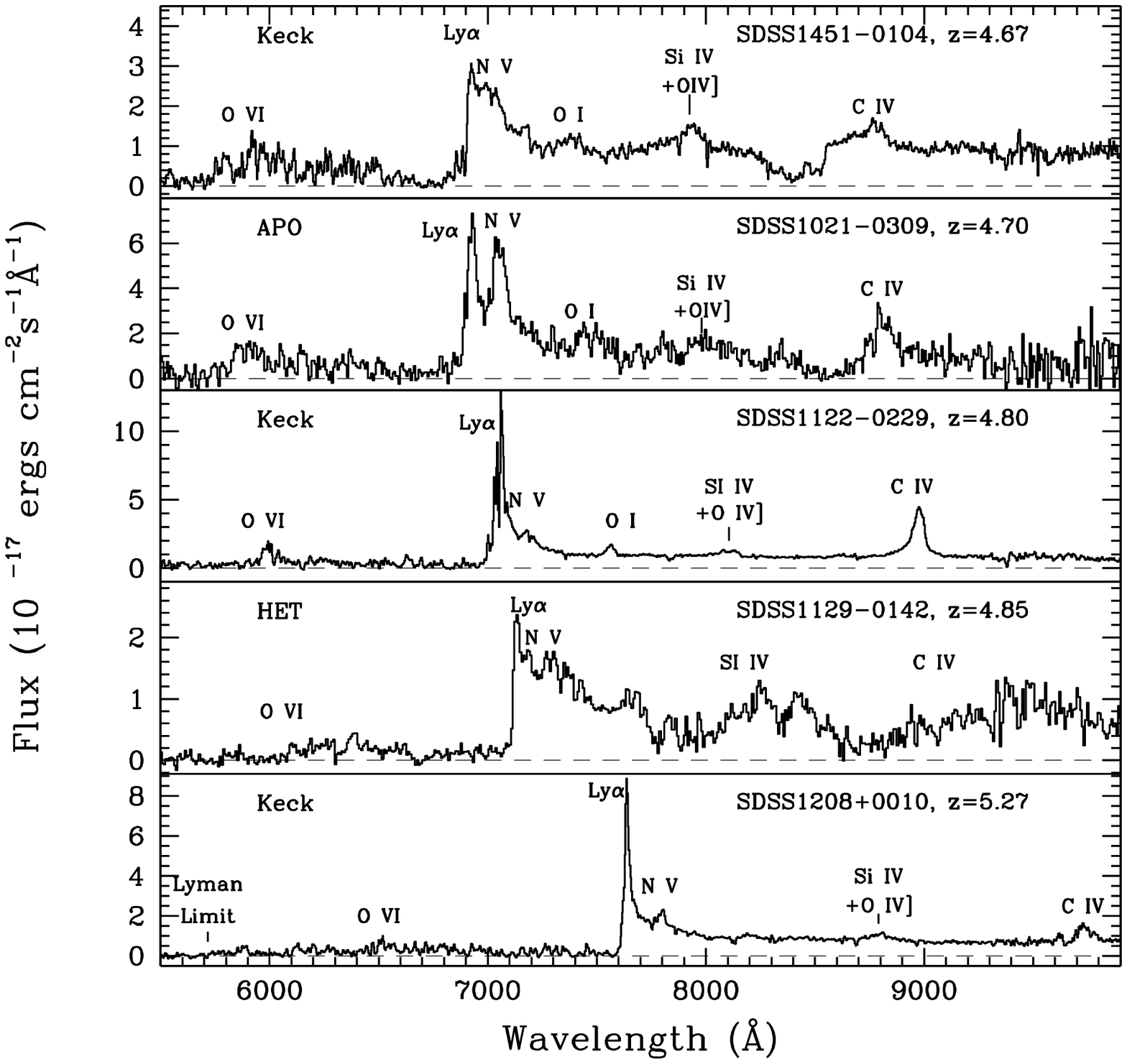]{Spectra of the five quasars. The wavelengths of
identified and expected major emission features are marked. The telescope
names for individual spectra are marked at the upper-left corner of each
panel.
The fluxes have been normalized to the respective SDSS $i^*$-band
magnitudes.
The spectral resolution is 13\AA\ for the APO data, 20~\AA\ for the HET
data,
and Keck spectra have been rebinned to 3.85~\AA.
\label{fig2}}

\newpage
\begin{footnotesize}

\begin{deluxetable}{ccccccccc}
\tablenum{1}
\tablecolumns{9}
\tablecaption{Optical Positions, SDSS\tablenotemark{a}\ \ and
Infrared\tablenotemark{b}\ \ Photometry}
\tablehead{
Object\tablenotemark{c} & $u^*$ & $g^*$ & $r^*$ &
$i^*$ & $z^*$ & $J$ & $K$ & E(B-V) 
}
\startdata
SDSSp J102119.16$-$030937.2
& 23.72 & 25.82 & 21.76 & 20.09 & 20.02 & 18.77 & 17.08 & 0.042 \nl 
& $\pm 0.55$ & $\pm 0.47$ & $\pm 0.09$ & $\pm 0.03$ & $\pm 0.10$ &
$\pm 0.07$ & $\pm 0.05$ & \nl
SDSSp J112242.98$-$022905.1
& 23.73 & 24.77 &22.22 & 20.38& 20.47& $\geq 19.5$& --- & 0.055 \nl
& $\pm 0.57$ & $\pm 0.50$ & $\pm 0.12$ & $\pm 0.04$ & $\pm 0.15$ &
$\pm 0.10$ & --- & \nl
SDSSp J112956.10$-$014212.4
& 24.06 & 25.26 & 22.02 & 19.64 & 19.51 & 17.51 & 16.09 & 0.072 \nl 
& $\pm 0.55$ & $\pm 0.37$ & $\pm 0.10$ & $\pm 0.03$ & $\pm 0.07$ &
$\pm 0.05$ & $\pm 0.05$ & \nl
SDSSp J120823.82+001027.7
& 24.39 & 24.77 & 22.75 & 20.79 & 20.72 & 19.43 & 18.10 & 0.024 \nl 
& $\pm 0.65$ & $\pm 0.31$ & $\pm 0.08$ & $\pm 0.27$ & $\pm 0.35$ &
$\pm 0.10$ & $\pm 0.10$ & \nl
SDSSp J145118.77$-$010446.2
& 24.40 & 24.45 & 22.65 & 20.70 & 20.53 & 19.43 & 18.17 & 0.044\nl
& $\pm 0.45$ & $\pm 0.47$ & $\pm 0.16$ & $\pm 0.04$ & $\pm 0.13$ &
$\pm 0.10$ & $\pm 0.10$ & \nl \hline
\enddata
\tablenotetext{a}{
Preliminary SDSS calibration; values are expressed as asinh magnitudes
(\cite{lupton99}).  The asinh magnitudes for zero flux in the
$u^*$, $g^*$, $r^*$, $i^*$, and $z^*$ bands are approximately
24.00, 24.85, 24.55, 24.20, and 22.40, respectively.}
\smallskip
\tablenotetext{b}{In UKIRT (Vega-based) magnitude system.}
\tablenotetext{c}{The J2000 coordinates are given in the object
name (hhmmss.ss+ddmmss.s).}
\end{deluxetable}

\begin{deluxetable}{ccccccc}
\tablenum{2}
\tablecolumns{7}
\tablecaption{Quasar Properties}
\tablehead{
Object & $z$ & AB$_{1450}$ & $\alpha$\tablenotemark{a} & $D_A$ & $D_B$ &
$M_B$\tablenotemark{b}
}
\startdata
SDSSp J102119.16$-$030937.2
& 4.696 $\pm$ 0.004 & 20.25 & $-1.6$ & 0.58 & 0.76 & $-26.3$ \nl
SDSSp J112242.98$-$022905.1
& 4.795 $\pm$ 0.004 & 20.58 & $-1.2$ & 0.64 & 0.79 & $-26.0$ \nl
SDSSp J112956.10$-$014212.4
& 4.85~ $\pm$ 0.03 & 19.22 & $-1.3$ & 0.84 & 0.97 & $-27.4$ \nl
SDSSp J120823.82+001027.7
& 5.273 $\pm$ 0.004 & 20.47 & $-0.7$ & 0.71 & 0.81 & $-26.3$ \nl
SDSSp J145118.77$-$010446.2
& 4.672 $\pm$ 0.004 & 20.42 & $-0.7$ & 0.71 &  0.86 & $-26.2$ \nl \hline
\enddata
\tablenotetext{a}{
Spectral energy index ($f_\nu \propto \nu^{\alpha}$); typical
uncertainties
are several tenths.}
\tablenotetext{b}{Calculated assuming $H_0$ = 50, $q_0$ = 0.5, and
a spectral energy index between the ultraviolet and blue band
of $-0.5$.}
\end{deluxetable}
\end{footnotesize}
\clearpage

\setcounter{figure}{0}
\newpage

\begin{figure}
\plotfiddle{zheng.fig1.ps}{8 in}
{0}{60}{60}{-152}{70} 
\caption{~}
\end{figure}

\begin{figure}
\plotfiddle{zheng.fig2.ps}{6 in}
{0}{90}{90}{-275}{-165}
\caption{~}
\end{figure}

\end{document}